# Vaccination and Complex Social Dynamics

Enys Mones[1], Arkadiusz Stopczynski[1,2], Alex 'Sandy' Pentland[2], Nathaniel Hupert[3] & Sune Lehmann[1,4]

[1] *Department of Applied Mathematics and Computer Science, Technical University of Denmark, Kgs. Lyngby, Denmark*

[2] *Media Lab, Massachusetts Institute of Technology, Cambridge, MA, USA*

[3] *Weill Cornell Medical College, Cornell University, Ithaca, NY, USA*

[4] *The Niels Bohr Institute, University of Copenhagen, Copenhagen, Denmark*

**Vaccination and outbreak monitoring are essential tools for preventing and minimizing outbreaks of infectious diseases[1–8]. Targeted strategies, where the individuals most important for monitoring or preventing outbreaks are selected for intervention, offer a possibility to significantly improve these measures[5,7,9]. Although targeted strategies carry a strong potential, identifying optimal target groups remains a challenge[5,10]. Here we consider the problem of identifying target groups based on digital communication networks (telecommunication, online social media) in order to predict and contain an infectious disease spreading on a real-world person-to-person network of more than 500 individuals. We show that target groups for efficient outbreak monitoring can be determined based on both telecommunication and online social network information. In case of vaccination the information regarding the digital communication networks improves the efficacy for short-range disease transmissions but, surprisingly, performance is severely reduced in the case of long-range transmission. These**

**results are robust with respect to the strategy used to identify targeted individuals and time-gap between identification of targets and the intervention. Thus, we demonstrate that data available from telecommunication and online social networks can greatly improve epidemic control measures, but it is important to consider the details of the pathogen spreading mechanism when such policies are applied.**

Strategies for controlling and containing various infectious diseases have been actively developed in recent years, the two most prominent methods being targeted monitoring and vaccination[5,11,12]. While full knowledge of the outbreak state (in case of monitoring and prediction) or full coverage (in case of vaccination) are preferable, acquiring information about every individual in the population of interest is rarely feasible[5,6]. For this reason *targeted* strategies that involve interventions focused on a small, carefully selected subpopulation have been recently considered[5,7,9,11,13]. It has been shown that densely-connected populations, such as schools[14,15], universities[5], or hospitals[16], play a major role in large outbreaks[17], offering numerous paths for diseases to propagate through society. At the same time, such cohesive communities can be used as a 'petri dish' for studying the structure of internal contacts, making them especially interesting in epidemiological efforts. In case of person-to-person transmission, direct identification of optimal target groups requires knowledge of the structure of the physical contacts network, collection of which is usually a time-consuming and complex task[5,14,18], limiting the feasibility of this approach at scale. Communication networks, such as online social networks or call detail records (CDRs) provide an accessible proxy of the structure of contacts among individuals, and immunization strategies that take advantage of the structure of these networks have been suggested[12]. Due to the known topological differences be-tween communication networks and networks of person-to-person proximity contacts[14, 18, 19], however, the value of digital communication networks for locating epidemiologically-relevant target individuals is not yet understood.

Here we explore the extent to which we can utilize the structure of communication networks to create effective epidemic interventions. We study a densely-connected population of 532 university students, based on data from the Copenhagen Network Study (CNS), which includes records of Facebook activity, call detail records, and Bluetooth scans collected with high temporal resolution over two years (details of the measurement are provided in Methods and in Ref.[18]. Using data from multiple layers of social interactions captured in the CNS study, we show that communication networks can be used to predict which individuals are central in the person-to-person network and are thus prime candidates for limiting epidemic spreading that depends on physical proximity. Further, our results show that the efficacy of counter-measures based on communication network target groups is strongly affected by the range of transmission: diseases that require very close encounters (similar to droplet spreading) can be effectively contained, whereas in case of long-range transmissions (closer to airborne diseases) target individuals found using communication networks play a less central role during an outbreak. We find that strategies based on communication networks are robust with respect to temporal changes, that is, target groups can be identified months before an epidemic outbreak and still provide a significant improvement over interventions based on randomly selected groups, provided that a disease spreads via short-range transmission.

Our analysis of physical proximity between individuals is based on two distinct networks which we call *full-* and *short-range* networks. These two networks correspond to contact sequences of two physical proximity ranges, which can be regarded to be rough approximations for airborne and droplet modes of pathogen transmission[20–22]. The full-range network contains all physical proximity interactions between participants up to the full Bluetooth range ($\sim$ 10-15 meters), while short-range network is constrained to approximately 1 meter, based on received signal strength indicator (RSSI)[22,23]. Communication networks are created from Facebook interactions and CDRs. All networks describe interactions within a single population of 532 participants over a period of one month (see Methods). The temporal resolution of the physical proximity data is 5 minutes, where a proximity interaction is included in a time binned network when the Bluetooth sensor registers a contact in the corresponding period. The 5 minutes time scale has been shown to capture the key dynamics of the person-to-person network[14,24]. Here we show that it is possible to extract information based solely on communication networks (Facebook and call networks) and to apply that knowledge in a fundamentally different system, i.e., the network of physical interactions supporting infectious disease spreading.

While activity in all four networks (full- and short-range proximity, Facebook, and call networks) follows distinct daily schedules and circadian rhythm, the maximum increase of intensity in digital channels occurs outside of work periods: lunch breaks, evenings, and weekends (Fig. 1a). Both full- and short-range networks of physical proximity feature remarkably higher edge density compared to social networks (Fig. 1b, c). The degree distributions of the person-to-person networks are consistent with an approximate normal distribution, whereas those of the online social networks follow a power-law distribution, supported by both the Akaike information criteria and Kolmogorov–Smirnov goodness-of-fit based model selection, confirming earlier findings[14,18]. Basic structural characteristics of the static aggregated networks are shown in Table 1.

The proximity networks contain a large number of links (in the order of $10^5$-$10^6$ across the observation period), revealing a highly dynamic network of real world contacts, in contrast to the digital communication networks exhibiting sparser and slower network dynamics. Figure 1d shows how these differences are reflected in the time respecting network connectivity, where the majority of nodes in the person-to-person networks can be reached in a relatively short time—over 40% in a day and over 95% in under a week—contrary to the digital communication networks which require more than a month to be fully explored, as quantified by the invasion percolation process with transmission probability 1 (Fig. 1d), i.e., a process that propagates from node to node across every edge of the network without fail. When compared to the full-range network, the short-range network includes only the most frequent contacts, suggesting a social tie[22,23], as quantified by the slope of the invasion curve: after a natural delay in the invasion level ($\sim$10 hours), a majority of the giant component in the short-range network can be explored in a shorter time, meaning that most of the links describe frequent contacts in that network (black dashed lines are included to highlight the slopes of the short- and long-range contact networks). The short-range network therefore resembles social channels more closely than the full-range network and we expect epidemic counter-measures based on the structure of communication networks to be more efficient in case of short-range transmission than that in the full-range case. The distributions of contact durations and waiting times (i.e., duration of time between two consecutive appearance of a contact) in the proximity networks are in agreement with previous findings in the literature, showing long-tailed distributions[25].

The robustness of the aggregated networks can be illustrated by the change in the size of their giant component (i.e., the largest connected component in a graph) after the removal of a random set of links as shown in Fig. 1e. When removing $f_{\text{edge}}$ fraction of the links randomly from the network, physical proximity networks break down at a density close to that of the random networks, whereas social networks become disconnected when a smaller fraction of links are removed (Fig. 1e). The findings above are consistent with previous work[5,7,14]: proximity networks are structurally homogeneous with a well-defined average degree, while communication networks are characterized by heterogeneous degree distributions[19,26]. In summary, physical proximity networks—which approximate the actual paths supporting the spreading of infectious diseases—and communication networks are fundamentally different both in a structural and dynamical sense, e.g., how infectious diseases spread through these networks.

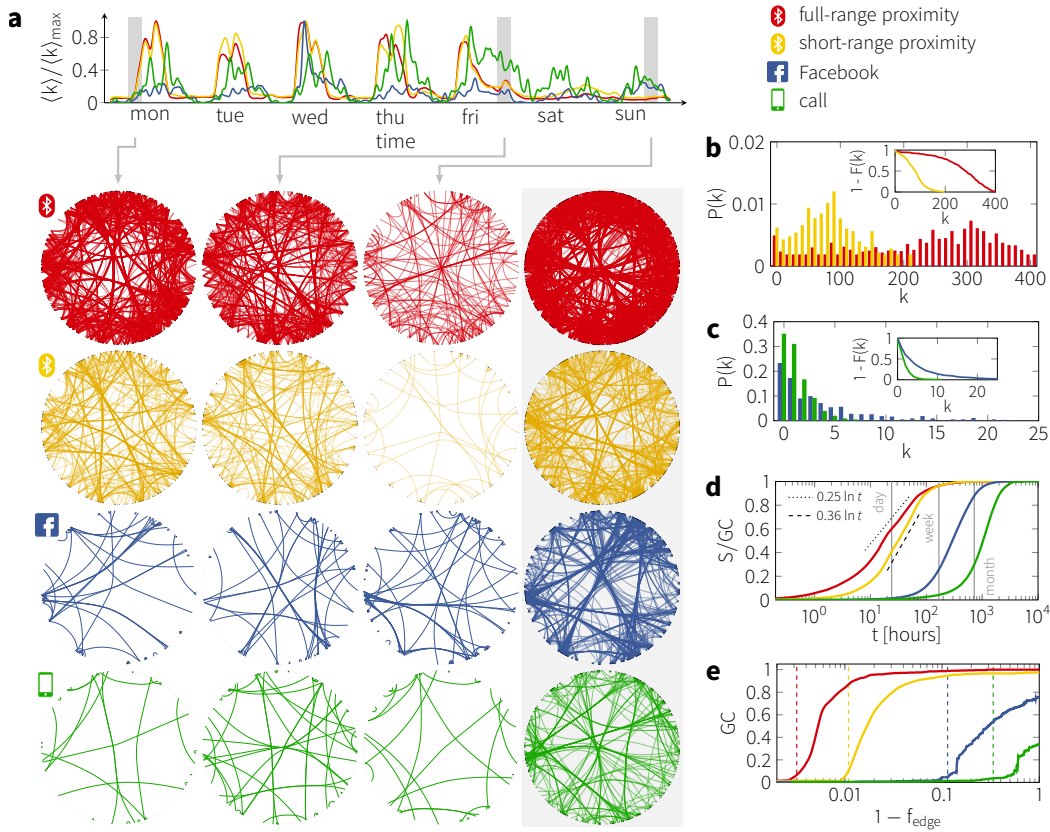

**Figure 1 Characteristics of the contact and communication networks.** Properties of the four networks aggregated over a week: person-to-person networks (red: full-range, yellow: short-range), Facebook (blue) and call (green) networks. For three specific times of the week, 4-hour snapshots are shown. The last column shows the graphs aggregated over the complete period of the analysis. **a)** Snapshots of the three networks aggregated over 4 hours periods: Time series display the corresponding dynamics of average degrees. Throughout the paper, information extracted from the communication networks (blue and green) is applied for intervention during an epidemic taking place on the person-to-person networks (red and yellow). **b)** Degree distribution of the proximity networks. Inset shows the corresponding cumulative distribution. **c)** Degree distribution of the communication networks. Inset shows the corrensponding cumulative distribution. **d)** Fraction of the giant component (S), i.e., the largest connected component, reached in an invasive percolation process with percolation probability equal to 1. For each temporal network, curves show the size of the percolation cluster relative to the giant component, the black lines are guidance for the slopes. All curves are averaged over 1000 random realizations of time and location.

**e)** Robustness of the aggregated networks against link removal. Giant component is shown after removing $f_{\text{edge}}$ fraction of the strongest links. Dashed lines indicate the percolation threshold of the corresponding random network with the same average number of first and second neighbors.

Below we explore how communication networks can be used for targeted intervention designed to observe and limit disease spread in person-to-person proximity networks. Targets for epidemic monitoring and vaccination are selected based on closeness centrality in the aggregated communication networks, i.e., the average distance of a node to all other nodes. Closeness centrality is chosen because this measure provides the best performance on our dataset. We are interested in the dynamics of a single epidemic event, therefore we focus on epidemics without an endemic state. We evaluate the efficacy of the counter-measures focusing on specific target groups by measuring the relative outbreak size during Susceptible-Infectious-Recovered (SIR) epidemics simulated on both the short and full-range proximity networks. Epidemic parameters are chosen to be consistent with expected infectious periods (3-4 days) and basic reproduction number ($2 < R_0 < 3$) of real-world infectious diseases such as influenza[27]; for details see Methods and the Supplementary Information. Communication network target groups are compared to *random control* and *colocation* target groups. In random selection, a number of individuals are chosen randomly from the population, providing a lower bound for intervention performance; conversely colocation target groups include individuals with highest fraction of time spent in proximity of others. Colocation-based target groups rely on full knowledge of the person-to-person proximity network topology, and while not strictly optimal, as they do not consider temporal dynamics, they establish an approximate upper bound for intervention performance[4].

In targeted monitoring, the objective is to select a small fraction of the population that is expected to become infected at an earlier stage and with a higher probability than the population average. In the monitoring scenario no vaccination is applied. Figure 2a depicts the median infection time restricted to the social target groups (i.e., individuals selected based on closeness centrality in digital communication network) compared to randomly selected and colocation-based groups. As a measure of performance, we use *corrected time of infection*, calculated as the average time the node is infected, normalized by the probability of infection across many epidemic realizations[4]. As expected, colocation-based groups display a significant lead time, a shift in the expected time of infection compared to the population average, showing a peak in the fraction infected before the epidemic peaks in the population[4,5]. We find that individuals selected using only communication channels also show improved lead times (24% and 18% compared to population average, corresponding to 3 and 2 days) for both short- and long-range networks–only 14% and 19% worse than the optimal (but logistically infeasible) colocation strategy, respectively.

The aim of vaccination is to decrease outbreak size by decreasing the number of links between individuals in a population, across which the disease may be transmitted. Therefore, the goal of an optimized targeted vaccination is to choose individuals with the greatest likelihood of infecting the largest number of their network neighbors. Specifically, a good candidate for immunization should be highly exposed to the disease and simultaneously exhibit high potential to transmit the infection. To assess the efficacy of vaccination based on the digital communication networks (social vaccination), we measure the relative epidemic outbreak size in the presence of immunized target group; that is, the total number of infections ($I_\infty$) divided by the initial number of susceptible ($S_0$): $i_{\text{rel}} = I_\infty/S_0$, excluding the vaccinated subpopulation from the calculation, and thus measuring only the network effect. In Figure 2b, the median relative outbreak sizes are plotted against the fraction of the population immunized. Using complete information available in the person-to-person network (the *hypothetical* colocation strategy), it is possible to reduce the outbreak size by more than $80\%$ after the immunization of $20\%$ and $27\%$ of the network in short- and full-range transmission respectively, effectively containing the outbreak.

In contrast to the general relationship just described between outbreak parameters and monitoring strategy based on proximity, the efficacy of social vaccination strongly depends on the nature of pathogen transmission, specifically its range. For diseases spreading via short-range interactions, Facebook and CDR based strategies are effective and outperform random immunization even for small target groups, approaching the performance of the colocation strategy once more than $20\%$ of the population has been vaccinated. If the disease transmission occurs on the full-range person-to-person network, however, communication networks do not significantly outperform random vaccination. This effect is more pronounced in the small target group size regime (less than $20\%$ of the population). Figure 2b reveals an inherent difference between the fraction of immunized individuals needed to reach the same reduction of outbreak size in the short and full-range networks, respectively. The effect of vaccination is generally weaker in the highly connected full-range network. To account for this difference, insets show the difference of outbreak size between communication network target groups and optimal groups for the same group sizes ($\Delta i$) as a function of outbreak size during optimal vaccination. The separation of curves indicates the difference in vaccination performance beyond the difference due to network connectedness. At low levels of immunization ($f_v < 0.1$, subplots (i) and (iv) in Fig. 2b), the effect of vaccination is low, due to the high density of edges in both person-to-person proximity networks. At high levels of immunization (Fig. 2b subplots (iii) and (vi)), with more than half of the population vaccinated, the digital communication network target groups contain the majority of socially active individuals, resulting in the social vaccination reaching the optimal strategy also for full-range transmission. In the intermediate range ($0.1 < f_v < 0.5$, subplots (ii) and (v) in Fig. 2b), communication network vaccination is significantly more effective and approaches optimal for the short-range transmission. The intuitive reason behind this difference between the short and full-range networks is that the structure of the full-range proximity network is strongly influenced by many random encounters, not captured by the social networks (see Supporting Information for full details).

The performance of social monitoring and vaccination is robust with respect to different centralities used for selection of the target groups (degree, k-coreness, betweenness) and variation in how communication channels are constructed. See Supporting Information for full details regarding the robustness of the results.

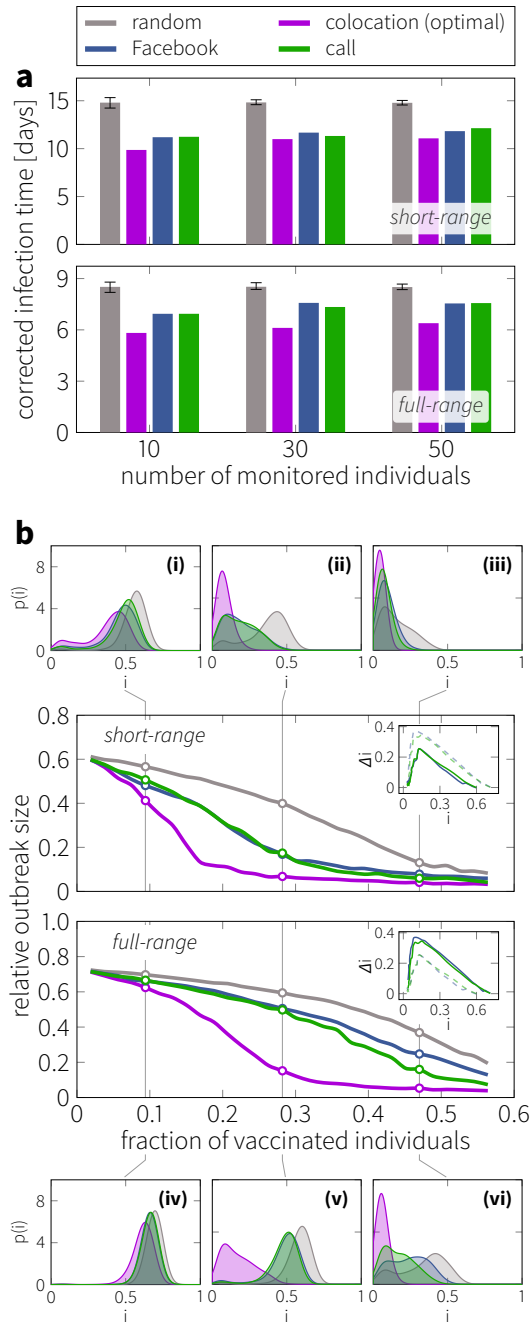

**Figure 2  Monitoring and vaccination efficiency of target groups. a)** Median corrected infection times among members of the social target groups compared to random (grey) and optimal (purple) groups in the absence of vaccination. Corrected infection time is the average infection

time normalized by the probability of infection for each individual. Results are based on 10 000 simulations. In case of random target groups, curves show the median of average infection times over 1000 random groups, and error bands denote lower and upper quartiles of the averages. **b)** Relative outbreak sizes (total number of infected divided by the initial number of susceptible) at different levels of immunization based on the social channels during short-range (top) and full-range (bottom) transmission. All curves show the median of 1000 simulations with minimum outbreak sizes of 5% among the initial susceptible. For three target group sizes (10, 30 and 50), the corresponding outbreak size distributions $p(i)$ are also shown. Insets show the difference $\Delta i$ of outbreak sizes between social vaccination and optimal, as a function of optimal outbreak size, $i$. Each inset shows all four comparisons, curves corresponding to the type of transmission in main plot (short or full-range) are highlighted.

Practical targeted monitoring and vaccination, regardless of the target selection strategy, requires the social system to be stable, so that targets can be identified before an outbreak takes place. To analyze the tradeoff between performance of targeted intervention and the time gap between identification and intervention, we fix the period that forms the basis of our target individual selection (February, *index month*) and calculate monitoring and immunization performance for epidemic outbreaks in subsequent months (March, April and May, *outbreak months*), as shown in Figure 3. The performance of strategies based on the index month is compared to random and two types of colocation strategies: colocation-based on proximity in the index month and that of the outbreak month. Communication network target groups show significantly lower infection time relative to random monitoring, although target groups based on the index month do not perform as well as the optimal groups calculated in the outbreak months (Figure 3a) due to the small changes in the social structure of the population. Immunizing members of the index month target groups outperforms random vaccination in all three outbreak months, as seen in Figure 3b. Further, relative outbreak size during social vaccination approaches those corresponding to the index month colocation strategy. These results suggest that communication networks enable us to locate individuals who consistently play a central role in epidemic outbreaks.

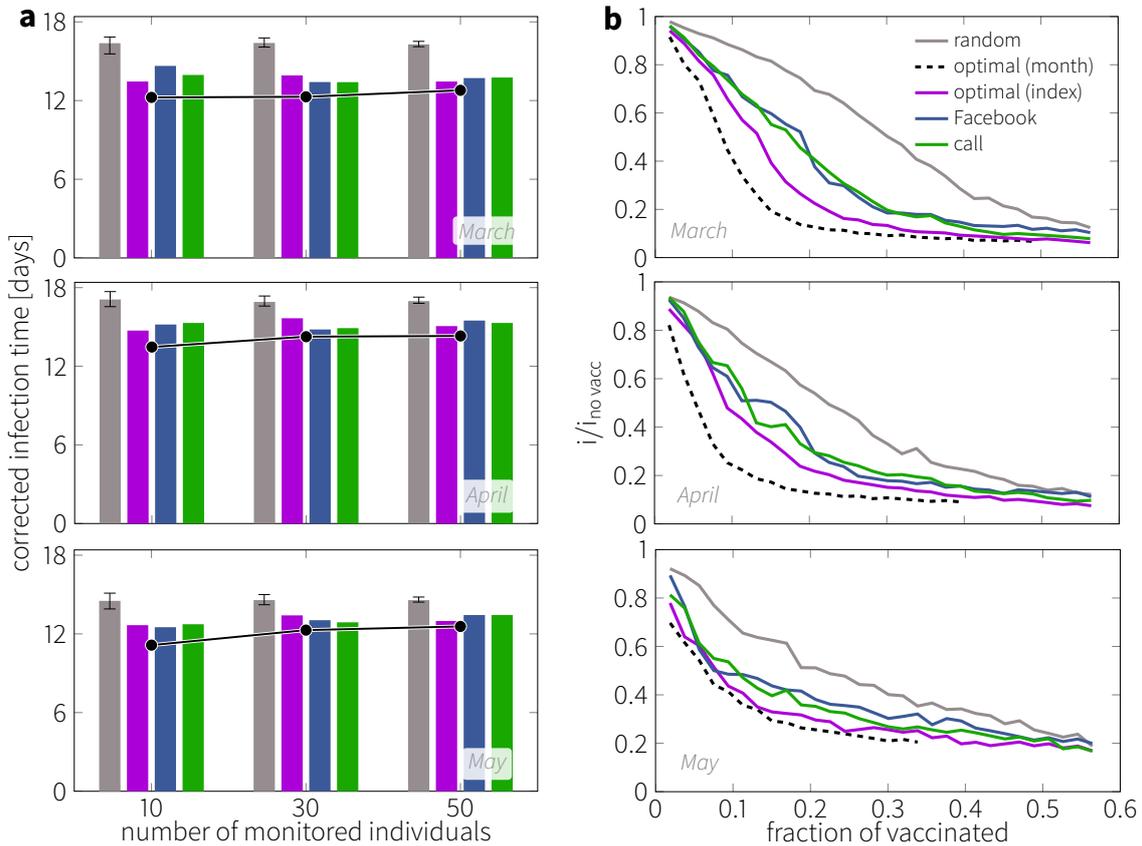

**Figure 3 Robustness of social target group effectiveness. a)** Median infection time as a measure for monitoring effectiveness in consecutive months: social (green and blue for call and Facebook based communication groups), optimal based on the index month of February (purple), random (grey) and optimal based on the monthly data (black line). Error bars on the random monitor groups correspond to lower and upper quartiles. **b)** Performance of social vaccination based on the index month applied in consecutive months. Communication network based vaccination is compared against random (grey), optimal for the index month (purple) and optimal for the corresponding month (dashed dashed). Curves represent median values over 1000 simulations.

In case of epidemic outbreaks, person-to-person proximity contacts provide the paths for many infectious diseases to spread, but measuring these contacts in vivo during an epidemic event is not feasible using current technology. In this paper, we have shown that the structure of digital communication networks does indeed allow us to identify candidates for targeted epidemic monitoring and vaccination. Monitoring based on communication networks outperforms random target groups and approaches the optimal strategy. The performance of targeted vaccination based on communication network structure, however, is strongly affected by the nature of pathogen transmission, displaying high efficacy in case of short-range transmission, but is significantly less applicable when an infection spreads via long-range contacts. This result arises from the inherent structure of short-range physical contacts: close contacts in the person-to-person network frequently correspond to social ties and therefore communication networks contain more relevant information about the structure of the short-range network. These findings suggest that, when considering real-world diseases, online social networks and CDR data can serve as a valuable resource for epidemic intervention. Regarding one of the most basic differences between the above transmission types (their physical range of infection) immunization strategies may benefit from digital communication networks in case of droplet diseases, but will have limited utility during airborne infections, as we expect real-world airborne networks to have even more connections between socially not connected individuals than the long-range network examined here due to the characteristics of airborne diseases (the ability to suspend in rooms, transmission via touches, etc). Below, we discuss and address the potential shortcomings of the dataset as well as the epidemic model used in the paper.

*Spreading mechanism*—It is important to acknowledge that our model of epidemic processes on the person-to-person proximity networks is a strong simplification of the underlying biological processes. The detailed transmission of droplet and airborne diseases is not fully captured by reciprocal Bluetooth measurements, e.g., transmission of biological pathogens is not merely characterized by the distance but is also affected by many other environmental characteristics and individual behavior. Droplet transmission requires individuals to face each other in close proximity[20,28], while airborne pathogens can stay suspended in the air or settle on surfaces, significantly increasing opportunity for infection[28]. However, these characteristics amplify the differences between the person-to-person networks by effectively removing superfluous links from the short-range network and adding additional noise to the long-range network. Thus, the efficacy of using digital communication channels in targeted monitoring and vaccination can be expected to be even higher for true droplet network and closer to random for airborne transmission.

*Optimal strategies*—Our strategy is based on basic network topology measures (centralities), and presumably more sophisticated strategies can be constructed in order to achieve higher performance in much larger populations. Our results, which suggest that social monitoring and vaccination are effective strategies to observe and limit epidemic outbreaks, are robust in the face of changes in the exact method of target selection (i.e., exact measure of centrality in the contact or communication networks), see Supporting Information for details.

*Epidemic modeling*—The participants in the study represent a single specific population, a fact known to result in an underestimation of epidemic risk[29]. Nevertheless, it should be noted that our results rely solely on the comparisons within the same population, and the structural differences are restricted to that of connectivity. In order to isolate the key mechanisms, we used a highly simplified model, the SIR dynamics, with no vital dynamics (temporal changes in the population due to births and deaths) or aging (accounting for different age classes of the population). It is well known that vital dynamics and aging have a strong impact on the spreading process[30], however, in the cohort that forms the basis of the current study, individuals are members of the same age class and therefore we can neglect vital dynamics. Finally, Bluetooth signals are able to pass through walls which introduces non-physical contacts and unrealistic spreading events in the dataset, however, the fraction of links that can be potentially associated to these cases is negligible.

In this paper, we have taken a first step towards investigating how knowledge of digital communication network structure of a population can be used to identify target groups for efficient targeted monitoring and vaccination. Our most notable finding is that, in our modeling framework, relatively subtle differences in the structure of transmission mechanisms, formulated here as a dichotomous choice between droplet-like and airborne routes, may have a profound impact on final epidemic size when using communication networks as a basis for targeted vaccination. These result may also have important practical implications for the distribution of non-vaccine epidemic countermeasures in these settings. Since digital communication data of the types modeled here may allow for early detection and containment of infectious outbreaks in densely-connected populations (i.e., in schools, universities, workplaces, and neighborhoods), our work supports increased collaboration between practitioners in public health and operators of social networks and telecoms.

**Methods**

**Data collection.** Proximity, call detail records (CDRs), and Facebook feed data was recorded during the Copenhagen Network Study between 2012 and 2014[18]. During the experiment, 1 000 Nexus smartphones with pre-installed data collector application were handed out to students at the Danish Technical University (DTU) and data on multiple channels was recorded: location, WiFi scans, Facebook feed, call detail records, Bluetooth scans with a temporal resolution of at least 5 min. Students used the devices as their phones, resulting in a high coverage ratio of the data during the period of the experiment; their anonymized data was uploaded to the servers at DTU. Results reported in this paper correspond to the period of February 2014. Here we consider data from the participants with at least 60% of data quality in the period of interest (based on proximity data), resulting in 532 individuals (proximity data was time-binned in 5 minutes bins, and users having Bluetooth signal or appearing in a Bluetooth scan list at least 60% of the total time bins were considered). See Supplementary Information for the details of data collection and data filtering.

**Ethical statement.** Prior to the experiment, all participating students were informed on the data collection method and the research goals. Data collection, anonymization, and storage were approved by the Danish Data Protection Agency, and comply with both local and EU regulations.

**Contact and communication channels.** Bluetooth scan lists are used to construct temporal proximity networks in two different ways: raw scan list data provides all interactions in a long-range (up to $10-15\,\mathrm{m}$) that are the basis for the full-range disease spread. By restricting contacts to signal strengths of $\mathrm{RSSI} > -75\,\mathrm{dBm}$, close-contact networks (up to $1\,\mathrm{m}$) can be obtained[23]. It should be noted that the above threshold is highly conservative in order to reduce the likelihood for observing long-range contacts, therefore the usual distance corresponding to the short-range network is below $1\,\mathrm{m}$. In the case of digital communication networks, both CDR and Facebook feed data were aggregated over the period of interest, resulting in undirected and weighted static graphs, where weights correspond to the number of interactions between participants.

**Epidemiological model.** Dynamics of the close- and long-range diseases is simulated using the SIR model. According to the model, all individuals are susceptible with the exception of a small number of index cases (1% in our simulations). Whenever an infected is in contact with a susceptible, the latter becomes infected with probability $\beta$, and at each time step each infected becomes recovered with probability $\gamma$, these parameters are adjusted to be consistent with real-world infections. The probability of infection per contact event and expected infectious period at each time step are $\beta^{\mathrm{full}} = 0.002$, $T_{\mathrm{inf}}^{\mathrm{full}} = 3\,\mathrm{days}$ and $\beta^{\mathrm{short}} = 0.01$, $T_{\mathrm{inf}}^{\mathrm{short}} = 4\,\mathrm{days}$ for the full- and short-range proximity, respectively. Considering the average degree in a time step, i.e., $\overline{\langle k \rangle} = \frac{1}{\mathcal{N}(t)} \int_{t_i}^{t_i + \Delta t} \langle k \rangle(t') \mathrm{d}t'$, (where $\mathcal{N}(t)$ denotes the number of time steps), the above infection probabilities correspond to real physical rates of infection of $\beta_{\mathrm{physical}}^{\mathrm{full}} = 0.717\,\mathrm{day}^{-1}$ and $\beta_{\mathrm{physical}}^{\mathrm{short}} = 0.591\,\mathrm{day}^{-1}$, and result in the basic reproduction numbers of $R_0^{\mathrm{full}} = 2.151$ and $R_0^{\mathrm{short}} = 2.364$, which is within the range of $R_0$ for influenza[11]. When individuals become infected, they recover with a probability that corresponds in the aforementioned average infectious times. More details on the parameter adjustment, analysis and the behavior of SIR dynamics on the presented proximity networks are presented in the Supplementary Information.

Throughout the paper, we applied several network theoretical measures that are well-defined in an undirected graph. Structure of a static aggregated network can be described by three basic network descriptors: the average number of contacts (average degree); average fraction of connected neighbors (average clustering coefficient); or the average number of steps between all pairs of nodes (average path length). Moreover, target groups of size $n$ are obtained from the digital social networks by ranking individuals in the aggregated graphs by their closeness centrality, defined by:

$$C_c^{(i)} = \frac{N-1}{\sum_{j \neq i} d_{ij}}, \tag{1}$$

where $N$ is the number of nodes in the graph and $d_{ij}$ denotes the geodesic distance between nodes $i$ and $j$, i.e., the lowest number of steps to reach node $j$ from node $i$. If the graph is not connected, it is defined to be $d_{ij} = N$, which is always larger than the geodesic distance between any pairs of nodes. After ranking the individuals according to $C_c$, we select the ones with the $n$ highest centrality. In case of colocation based target groups, individuals are ranked based on their total time spent in the proximity of others, that is:

$$w_i = \sum_{j,t} \gamma_{ijt}, \tag{2}$$

where $\gamma_{ijt} = 1$ if participants $i$ and $j$ have been in close proximity at time $t$, and zero otherwise. Ranking all members by their weight $w_i$, we select the ones with the largest value to include in the target groups, following the strategy of Smieszek and Salathé[4]. Strategies based on other centrality measures as well as the details of the selection are discussed in more details in the Supplementary

Information.

**Acknowledgements** We thank L. K. Hansen, and P. Sapiezynski, for invaluable discussions and comments on the manuscript. This work was supported a Young Investigator Grant from the Villum Foundation (High Resolution Networks, awarded to S.L.), and interdisciplinary UCPH 2016 grant (Social Fabric). Due to privacy implications we cannot share data but researchers are welcome to visit and work under our supervision.

**Competing Interests** The authors declare that they have no competing financial interests.

**Correspondence** Correspondence and requests for materials should be addressed to E.M. (email: enmo@dtu.dk).

| Network | $E$ | $\langle k \rangle$ | $C$ | $\ell$ |
|---|---:|---:|---:|---:|
| Full-range proximity | 69 055 | 259.6 | 0.644 | 1.53 |
| Short-range proximity | 20 690 | 77.78 | 0.356 | 1.91 |
| Facebook | 1 261 | 4.741 | 0.150 | 3.87 |
| Call | 354 | 1.331 | 0.102 | 7.03 |

Table 1: Structural properties of the static aggregated physical proximity and social networks: number of diads ($E$), average degree ($\langle k \rangle$), average clustering coefficient ($C$) and average path length ($\ell$). For definitions, see Methods.

# Vaccination and Complex Social Dynamics
## *Supplementary Information*

Enys Mones[1], Arkadiusz Stopczynski[1,2],
Alex 'Sandy' Pentland[2], Nathaniel Hupert[3] & Sune Lehmann[1,4]

March 2, 2016

1. Department of Applied Mathematics and Computer Science, Technical University of Denmark, Kgs. Lyngby, Denmark
2. Media Lab, Massachusetts Institute of Technology, Cambridge, MA, USA
3. Weill Cornell Medical College, Cornell University, Ithaca, NY, USA
4. The Niels Bohr Institute, University of Copenhagen, Copenhagen, Denmark

## Contents

## S1   The Dataset

We consider data collected in the Copenhagen Networks Study, spanning years between 2013 and 2015. The data has been collected from a densely-connected freshman population of approximately 1 000 students at a large European university (Technical University of Denmark) and contains high-resolution traces including close-proximity interactions, telecommunication, online social networks, and geographical location. Here, densely-connected refers to the high frequency of physical proximity contacts as well as online communication between the individuals. Majority of the data has been collected with custom-built application installed on smartphones provided to the participants (Google Nexus 4). Full details of the study can be found in Ref. [1].

| Network | $V_{\text{active}}$ | $E_{\text{dynamic}}$ | $E_{\text{static}}$ | $\langle k \rangle_{\text{dynamic}}$ | $\langle k \rangle_{\text{static}}$ |
|---|---|---|---|---|---|
| Full-range proximity | 532 | 2 670 547 | 69 055 | 1.245 | 259.6 |
| Short-range proximity | 518 | 428 481 | 20 690 | 0.200 | 77.78 |
| Facebook | 410 | 3 321 | 1 261 | 0.007 | 4.741 |
| Call | 345 | 2 134 | 354 | 0.004 | 1.331 |

Table 1: Basic statistics of the physical proximity and social networks: active individuals ($V_{\text{active}}$), those having at least one link; total number of temporal links ($E_{\text{dynamic}}$); number of diads or static links ($E_{\text{static}}$); dynamic degree ($\langle k \rangle_{\text{dynamic}}$), that is the average degree for single time bins, averaged over all time bins; static degree ($\langle k \rangle_{\text{static}}$). Average degrees are calculated using the total population size, i.e., $V_{\text{total}} = 532$.

In this manuscript we select individuals with a data quality of at least 60% during period we focus on (February 2014), defined by the time coverage of the Bluetooth scans. Data quality is calculated by binning all Bluetooth scans (that are used for inferring physical proximity interactions) and removing participants that have appeared in scans in less than 60% of the total bins (i.e., 4838 out of 8064 time bins). This results in a population of 532 individuals. Using Bluetooth, call detail records (CDRs), and Facebook feed data, we construct four distinct networks of interactions between the selected individuals.

## S1.1 Network creation

*Physical proximity networks* are built using Bluetooth scans between devices. Each time the device ID of participant $i$ is listed in a Bluetooth scan of the device of participant $j$ at time $t$, there is a link $\gamma_{ijt} = s$ between the corresponding participants, where $s$ denotes the received signal strength indicator (RSSI). From the raw counts, we build temporally-binned contacts with bin size of 5 minutes, such that if there is any observation in the $n$-th time bin, we add a link. As we do not expect false positives in a Bluetooth scan, all edges are undirected even if they are observed in only one direction. We also extend local star-like graphs to cliques, to avoid missing links in case of a single (or very few) devices reporting data in particular time bin. Such pre-processing results in physical proximity network capturing interactions at the distance of $10 - 15$ meters, which we denote *full-range* network. The network is described by a binary-valued adjacency matrix $A_{i \times j \times t}$ with $a_{ijt} = 1$ when interaction is present and $a_{ijt} = 0$ otherwise. We also create *short-range* network by thresholding the links based on their RSSI values, i.e., we restrict the links only to those for which $\gamma_{ijt} > s_0$ with $s_0 = -75$ dB. It has been shown that this value corresponds to a distance up to approximately 1m [2, 3]. Thus, we create two types of proximity networks, one with full-range interactions and one with only very close proximity events. This constraint on the signal strength results in a more sparse and tree-like network (18% of the original interactions are present after thresholding).

*Digital social networks*, corresponding to social behavior and social ties, are based on calls and Facebook activity. Each call between two individuals results in a corresponding link in the call network, and every interaction on Facebook creates a link in the *functional* Facebook network (interactions include comments, tags, messages, mentions, posts, etc). Here we do not use the temporal information in the social networks, i.e., we use networks as aggregated over the month of interest. We also do not consider directionality of calls nor directionality or type of Facebook interactions. It is also possible to build other types of networks by including text messages in addition to calls metadata or using the static friendship links between users (for *structural* Facebook graph). However, as we show in the sections Monitoring and Vaccination, these approaches do not result in significant differences in social monitoring and vaccination. Finally, we summarize the basic properties of the four networks of the study in Table 1.

## S1.2 Limitations of physical proximity contacts

As Bluetooth signals can pass through walls, some fraction of the contact networks built from the Bluetooth scans are expected to be false contacts, as they are not able to transmit real physical diseases. To address this problem, we consider the signal strength (RSSI) of the links: it's feasible to assume that links passing through walls have a vanishing strength irrespective of the actual distance. Therefore, in the short-range network we don't expect fake contacts as it includes only links with a certain high signal strength. In case of full-range network, weak links are more common. However, from the simulation perspective those weak links have a relatively high impact on the spread that are also frequently re-appearing, i.e., signals corresponding to dorm walls. To investigate these links, in Fig. 1 we plot the total time coverage of the links versus their average signal strength. Signals crossing dormitory walls are located in the upper-left corner, i.e., the frequent weak links, however, as the figure shows, these links are rare and they cover a negligible fraction of the total temporal links.

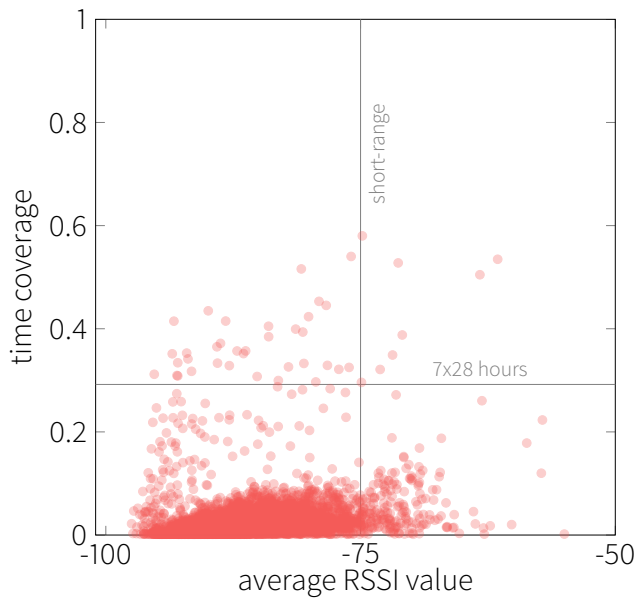

Figure 1: **Distribution of Bluetooth based proximity contacts** Total temporal coverage of proximity links versus their average RSSI, that is, the signal strength. Above the horizontal line are the links that describe contacts of frequent periods with a total length of more than $7 \times 28$ hours. To the left from the vertical lines lie the links that correspond to the short-range interactions ($RSSI > -75\,\text{dB}$).

Furthermore, the majority of infections take place during lunch breaks, classes and social events where individuals are located in the same closed area.

### S1.3 Target group selection

For the epidemic monitoring and vaccination, we construct target groups based on different types of ranking: random, colocation, and social.

In case of random target groups, for a given group size $n$, we pick $n$ participants from the population at random, without using any auxiliary information.

For colocation groups, we consider the aggregated weight of individuals in the proximity networks. For each participant we calculate $w_i = \sum_{j,t} a_{ijt}$, that is, the total time (expressed in 5-minute bins) spent in the proximity of others. Note that in this case, each temporal link is considered separately, thus someone spending a given amount of time in a large group has higher weight than that of spending the same time in a small group. After ranking individuals by their weight, we select the top $n$ individuals from the population as our target group. As shown in Ref. [4], in case of epidemic monitoring the colocation-based target group performs close to the optimal strategy achieved with a greedy algorithm. Although during vaccination the colocation-based group is not strictly optimal considering that no temporal information is utilized, it provides a reasonable upper bound of the vaccination performance with access to full information about close-proximity interactions. Keeping the above in mind, we refer to the colocation strategy as *optimal*.

Social target groups are based purely on information extracted from the social networks, with the assumption that those networks describe, at a fundamental level, social structure of the population. In order to assess the social role and importance of an individual, we calculate their centrality in the aggregated social networks, rank the population by the centrality, and pick the $n$ most central individuals. The results in the manuscript are reported with closeness centrality, however, the exact centrality used does not affect the findings qualitatively. Similarly, even if link weights (i.e., the number of contacts between individuals) are considered, the corresponding results do not change and we do not observe significant gain in performance, suggesting that the extent of the ego-network is more relevant than the strength of the specific links.

For some group sizes the ranking of the individuals does not result in distinct ranks, i.e., when selecting top $n$ participants is not unambiguous. In these cases we consider 10 target groups by randomly selecting the lowest rank members and average all results over the 10 samples. More precisely, let us assume we can select $n-k$ individuals with distinct rank and the remaining $k$ individuals share the same rank with $m-k$ others (in other words, the ones with the lowest rank in the target groups can be selected from among $m$ participants). Then we pick 10 different groups with the lowest $k$ members being sampled uniformly from the $m$ ones sharing the same rank, perform the same analysis and average the results over the groups.

## S1.4 Testing for quality bias

We perform a quality check on the data to ensure that the reported results are not simply a consequence of quality bias. We test if individuals with high-quality data (those having a high temporal coverage in Bluetooth scans) are also the ones that have many calls or increased activity on Facebook. As Bluetooth scanning happens passively on the phones and we consider data quality based on the number of scans (regardless whether they discovered other participants of the study or not), we expect no correlation between how socially active the participants are and their data quality. We plot call and Facebook event counts for each user versus their corresponding coverage of Bluetooth data, as shown in Fig. 2. For the comparison, we plot data calculated over a complete year and show both total and internal communications of participants. (for better visibility, we omitted the top $< 10$ largely deviating data points). Correlations are low and in many cases are not significant: $r(w, c_{\text{call}}) = -0.0343(0.4286)$, $r(w, c_{\text{call total}}) = 0.2012(0.0000)$, $r(w, c_{\text{Facebook}}) = 0.0009(0.9833)$, $r(w, c_{\text{Facebook total}}) = -0.0282(0.5149)$. We report Spearman correlations, not assuming linear relation between the intensity of the two activities (social and proximity), numbers in the parentheses denote the corresponding $p$-values. The results indicate that the image of social activity we get about the population is not driven by data quality variation. Note that the manuscript, we base our results only on internal phone calls.

As for the social channels, the likelihood of missing data in the Facebook channel is minimal as the feed of each user has been crawled every 24 hours, providing a high level of redundancy for data quality. We consider the quality of the call data instead. To investigate whether we perceive some students that are highly connected in the call network merely due to the quality of recording their events better than others', we compare the calling behavior pattern of all participants in the month of interest (February 2014) to their yearly behavior. Figure 3 shows the individual deviations from their yearly average in February compared to the call intensity in the index month, i.e., the standard score of monthly calls: $(n_i^0 - \bar{n}_i)/\sigma(n_i)$ versus $n_i^0$, where $n_i^0$ is the number of calls in the considered month, $\bar{n}_i$ and $\sigma(n_i)$ are the mean and standard deviation in the other months. As seen from the plot, high number of calls does not correlate with the deviation from the individual average. Furthermore, as centrality values (color of the dots) indicate, high social centrality is not driven by count number nor deviation from the yearly behavior. This indicates that although we may expect certain amount of noise in the data, relevant measures on the population are not biased.

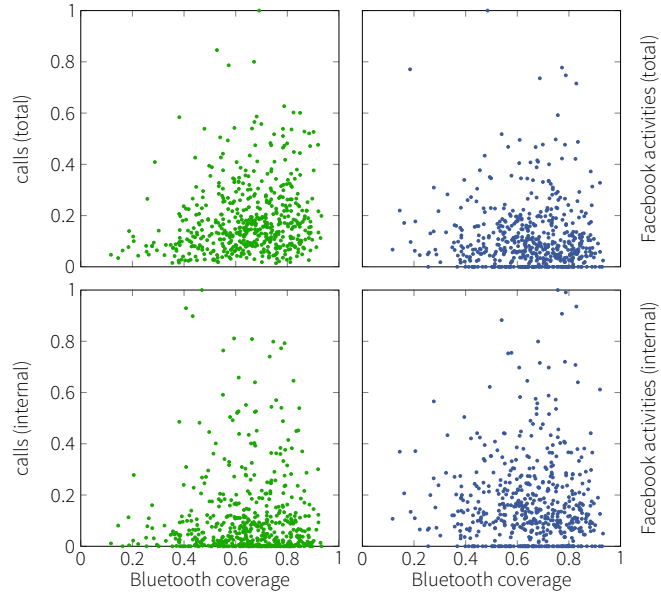

Figure 2: **Social activity versus data quality in the proximity layer** Call and Facebook activity intensity compared to the coverage seen by the Bluetooth scans for each participant in the current study: (top) total number of calls and Facebook events involving the given participant, (bottom) internal communication events, i.e., only between individuals considered in the simulations. Social network counts are normalized by the largest data point shown.

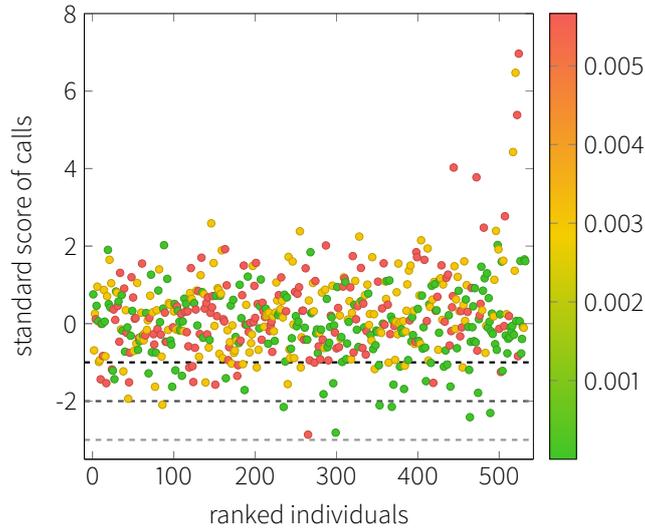

Figure 3: **Call activity deviations versus call counts** Standard score of monthly number of calls for each participant considered in the current study compared to their rank based on the actual number of calls in the index month (February, 2014). Color denotes the closeness centrality of the individuals in the aggregated network of calls.

## S2  Model

### S2.1  SIR model

For simulation of epidemic outbreaks, we use a susceptible-infected-recovered (SIR) model taking place on the network of physical interactions of the individuals. The SIR model assumes that the entire population—except for the index cases—is initially susceptible. Whenever a susceptible individual has contact to an infected one, the susceptible individual becomes infected with probability $\beta$. Infected individuals recover after an average infectious period $T_{\text{inf}}$ which is modeled in this study by a probability $\gamma$ to recover at any time. Recovered individuals can not be re-infected. We use the temporal proximity networks obtained from the Bluetooth scans as the contact network. In all simulations, the size of the index (initially infected) group is 5 (approximately 1% of the population).

### S2.2  Transmission types

We model two different types of disease transmission: full-range and short-range infections, by modifying the structural properties of the underlying networks as introduced in the previous section. In the full-range transmission (up to 10-15 meters) we loosely approximate spreading of airborne diseases (e.g. measles) [3]. Short-range transmission (approximately up to 1 meter) corresponds roughly to droplet diseases (e.g. influenza). While the proximity networks do not correspond to disease transmission exactly, they have been considered in the literature a telling approximation of the potential spreading paths [4, 5, 6, 3].

### S2.3  Parameter selection

The applied epidemiological model has two parameters: the probability of infection $\beta$ and the probability of recovery $\gamma$. Given the temporal resolution ($\Delta t = 1/288 \text{day}$) and temporal average degree ($\langle k \rangle = \frac{1}{\mathcal{N}(t)} \int_{t_i}^{t_i + \Delta t} \langle k \rangle(t') \mathrm{d}t'$, $\mathcal{N}(t)$ denoting the number of time bins), the simulation probabilities correspond to real physical rates in the following way. The rate of infection is given by

$$\beta_{\text{physical}} = \frac{\langle k \rangle \beta}{\Delta t}, \qquad (1)$$

whereas the expected infectious period is

$$T_{\text{inf}} = \frac{\Delta t}{\gamma}. \qquad (2)$$

Based on the physical rates the basic reproduction number is $R_0 = \beta_{\text{physical}} T_{\text{inf}}$.

Before investigating epidemic surveillance and immunization, we have to set infection probability and infectious time by bearing two aspects in mind: first, we intend to have an infectious time in the order of magnitude of days and second, to measure changes in outbreak sizes accurately we aim for an outbreak that lies near the boundary of the epidemic threshold: a disease that does not extinct but it does not result in the full infection of the population either. Figure 4 shows the relative outbreak size ($i_{\text{rel}}$, the total number of infected divided by the population size) at different values of the parameters during airborne and droplet diseases.

Note that parameter values were probed on a logarithmic scale (we tested parameter values of $10^n$, $2 \cdot 10^n$ and $5 \cdot 10^n$ for various exponents $n$) and therefore the size of the tiles in the figure is not uniform. Based on the outbreak size landscape, the following parameters were chosen for the airborne and droplet networks (marked by the white circles): $\beta^{\text{full}} = 0.002$, $T_{\text{inf}}^{\text{full}} = 3$ days and $\beta^{\text{short}} = 0.01$, $T_{\text{inf}}^{\text{short}} = 4$ days, resulting in typical outbreak sizes of $i_{\text{rel}}^{\text{full}} = 0.727$ and $i_{\text{rel}}^{\text{short}} = 0.634$. Table 2 summarizes the simulation parameters and related physical rates for the two proximity networks.

As we are interested only in the actual outbreak size (instead of the probability of outbreak or extinction), in all cases we accept only those simulations with a relative outbreak size of at least 5%, which results in a corresponding saturation value of the relative outbreak size in our vaccination curves. As for the monitoring, to obtain a larger ensemble for the statistics, the threshold of relative outbreak size is 20%, however, the results are robust against the choice of threshold.

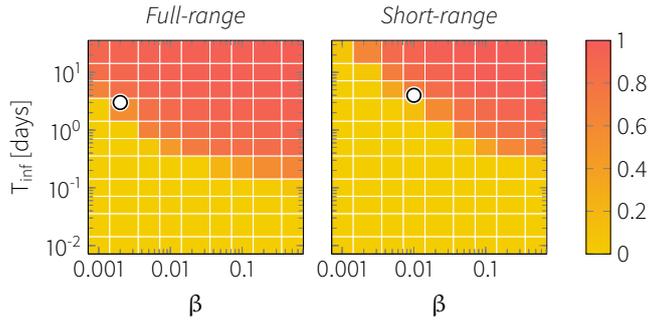

Figure 4: **Outbreak size as the function of the various paremters.** Absolute size of the outbreak (total number of infections normalized by the population size) at different values of the contact rate ($\beta$) and average infectious period ($T_{\text{inf}}$). White circles show the position of the actual values that were used for the two transmission types. Each tile corresponds to the median outbreak size of 100 SIR simulations.

| Network | $\langle k \rangle$ | $\beta$ | $\gamma$ | $\beta_{\text{physical}}$ | $T_{\text{inf}}$ | $R_0$ |
|---|---|---|---|---|---|---|
| Full-range | 1.245 | 0.002 | $11.57 \cdot 10^{-4}$ | $0.717\,\text{day}^{-1}$ | 3 day | 2.151 |
| Short-range | 0.200 | 0.01 | $8.68 \cdot 10^{-4}$ | $0.591\,\text{day}^{-1}$ | 4 day | 2.364 |

Table 2: Simulation parameters and the corresponding physical quantities: average temporal degree ($\langle k \rangle$), probability of infection for a single infected-susceptible contact ($\beta$), physical rate of infection ($\beta_{\text{physical}}$), probability of recovery for a single time step ($\gamma$), expected infectious period ($T_{\text{inf}}$) and the basic reproduction number ($R_0$).

### S2.4 Epidemic dynamics

Here we provide a quick overview of the actual dynamics of the epidemics and how it is realized in the various monitoring and vaccination scenarios. For the two cases of preventive strategies, we plot the fraction of infected in Fig. 5 in time. During epidemic surveillance, we consider a non-vaccinated scenario and measure the corresponding statistics inside the target group, which is shown in Fig. 5a. As the curves show, for both transmission types, the prevalence peaks at an earlier stage in social target groups compared to the population average. Furthermore, in case of the short-range network, the group dynamics is almost identical to that of the optimal group. The cumulative curves also indicate significant differences (Fig. 5a inset). In the presence of vaccination (Fig. 5b), we immunize those in the target group and plot the fraction of infected in the total population. For better comparison, we use a larger target group size (120 individuals) for the full-range network (for short-range, a group of 50 participants is shown), still we can observe large differences between the two cases of transmission types. Although social vaccination displays pronounced effects compared to the unvaccinated and the random vaccination cases, the slow-down in the dynamics and the the shrink of the peak is more emphasized during short-range interactions, also approaching the results of immunizing the optimal target group more. Again, the above effect is more visible in the integrated number of infected as shown in the inset of Fig. 5b.

## S3 Monitoring

In epidemic surveillance candidates in effective target groups are preferred to be in the high-risk subset of the population and preferably are infected early in the outbreak. This way, status of target individuals can be used to notice epidemic outbreak before it infects a high fraction of the population and may help in forecasting outbreak statistics. Therefore, the good target group is small compared to the population (so that monitoring is feasible and requires a lower level of resource allocation) and has high infection probability and low infection time. Here we consider various target group sizes and, in order to obtain a large number of samples for the statistics, we only consider outbreaks above 20% of relative outbreak size. In each case, we measure two quantities for target individuals related to epidemic surveillance. First we calculate the probability of infection—the fraction of simulations in which an individual becomes infected, provided that they are not part of the index group:

$$\mathcal{P}_{\text{inf}}(i) = \frac{N_{\text{inf}}(i)}{N(i)}, \qquad (3)$$

where $N(i)$ is the number of simulations in which individual $i$ was not an index case and $N_{\text{inf}}(i)$ is the number of simulations in which individual $i$ was infected during the outbreak. We also measure a corrected infection time that accounts for the level of exposure of an individual to the disease—the average time of their infection, divided by the probability of infection:

$$\tau_{\text{inf}}(i) = \frac{t_{\text{inf}}(i)}{\mathcal{P}_{\text{inf}}(i)}. \qquad (4)$$

where $t_{\text{inf}}(i)$ is the average time of getting the infection for individual $i$. In case of individuals who are never infected, the total duration of the outbreak is considered.

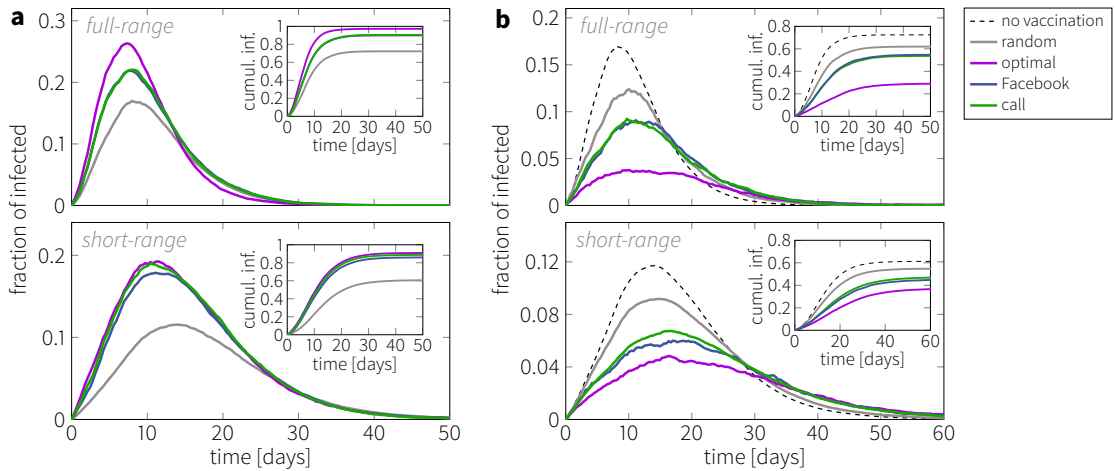

Figure 5: **Typical dynamics of the outbreak. a)** Prevalence of the disease inside the monitor group compared to random target groups (that shows the same dynamics as the population average). For each curve, monitor groups include the 30 individuals with the highest rank according to the given strategy. **a)** Prevalence measured in the total population during vaccination. Each curve represents the dynamics of the infection when the corresponding target groups are immunized, compared to the unvaccinated case and also to randomly selected target groups. Size of the target groups is 120 for full-range and 50 for short-range transmission. In both panels, insets show the cumulative fraction of infected. All curves are the median over 1000 simulations.

### S3.1 Social monitoring

Figure 6 shows the infection probability and infection time for the four types of target groups during full- and short-range infection outbreaks. In both transmission types, social monitor groups outperform random groups in the probability of infection, with performance close to optimal (colocation) groups, indicating that socially central individuals are indeed members of the high-risk subpopulation, irrespective of the infection range. Infection times in these groups are significantly lower compared to the population average, displaying temporal gain of 1.5 - 4 days, translating to leading times of 15-25% compared to the population mean, depending on the transmission type and group size.

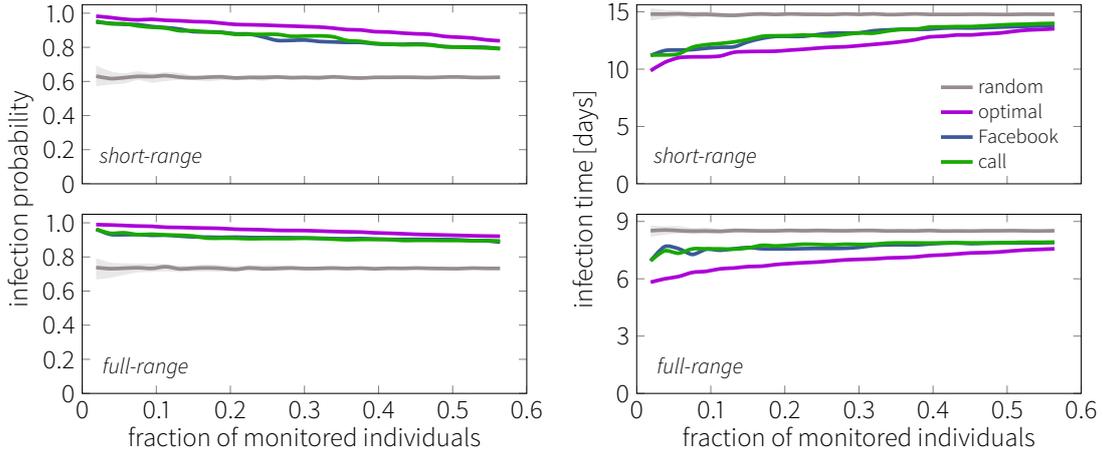

Figure 6: **Monitoring performance of social target groups** Median infection probability (left) and median infection time (right) of social target groups of different size, compared to the population average (random, grey) and optimal (purple) groups. Each curve is the result of $10^4$ simulations with a minimum relative outbreak size of 20% in an unvaccinated population. Curves corresponding to the random groups show the median over 100 random realizations, error bars denote lower and upper quartiles of the samples.

Social networks contain sufficient information about the population to effectively monitor the population in case of an epidemic, and various layers of social interaction (calls, text messages, Facebook events) show similar performance in inferring good candidates for monitoring. Furthermore, as long as the selection of target individuals is based on a centrality that grasps the global structure of the corresponding networks, the details of the method is less relevant. These observations are summarized in Fig. 7, where we compare various centralities as well as social network channels. Comparison of different centralities considered in the call-based network results in similar performance (with the exception of the raw count-based ranking), and we also notice that closeness centrality shows the best performance, although only marginally better than other strategies.

In addition to various centrality measures, we also consider social sensors strategy [7, 8, 9]. We simulate friendship based target groups from the phone network (calls and text messages) in the following way. First, we consider the three most frequent contacts for each individual and denote them as *friends*. Based on these top-contact lists, we rank the participants according to the number of times they appear on the friends list. As Fig. 7b shows, the resulting target groups have similar performance as the centrality-based ones. Finally, if additional or other channels as text messages or structural Facebook connections are considered, the monitoring performance does not change significantly (Fig. 7c-d).

## S3.2 Effect of network structure

Since monitoring does not alter the structure of the network but probes the outbreaks dynamics using small number of individuals, we can simplify the question of finding good candidates to possible correlations between network properties and monitoring efficiency, e.g., infection probability. We remove various correlations from the proximity networks, and plot infection probability versus the weight of individuals in the network. Results are shown in Fig. 8 for both full- and short-range proximity networks. We apply three types of correlation removal:

> *Temporal* We remove all circadian rhythm effects and re-distribute links homogeneously in time, keeping both aggregated degree-distribution and degree-correlations. We simply reassign time stamps so that their distribution became uniform in time. Individual weights are kept fixed.

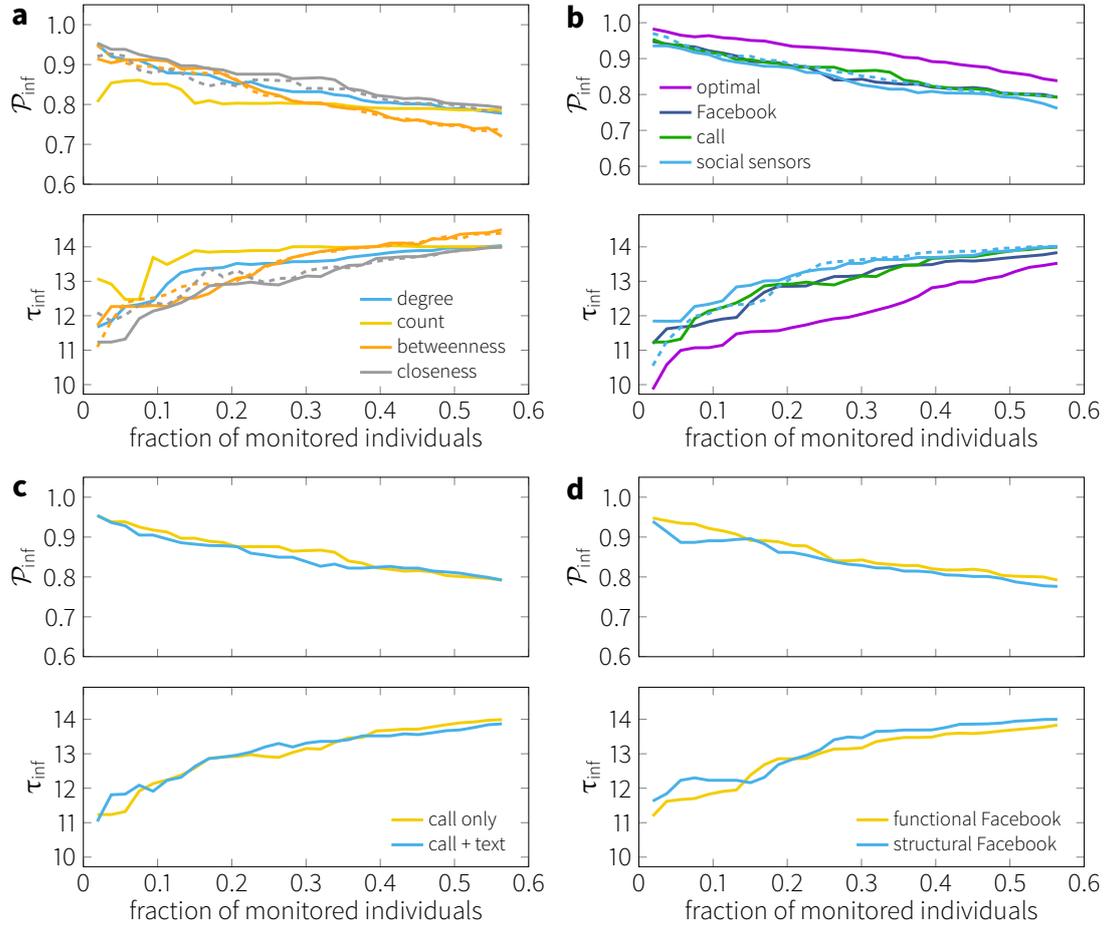

Figure 7: **Different monitoring strategies.** Monitoring power of target groups selected by various methods and using different channels: (a) testing centrality measures based on the call network, dashed lines denote weighted measures; (b) social sensors estimated from the call + text messages network as well as from the short-range proximity network, dashed lines correspond to social sensors estimated from the short-range proximity network; (c) networks constructed from calls only compared to calles + text messages; (d) functional Facebook network (based on user activity) compared to structural Facebook network (that is, the sole structure of friendships). All lines are the result of 1000 simulations with a relative outbreak size larger than 20%.

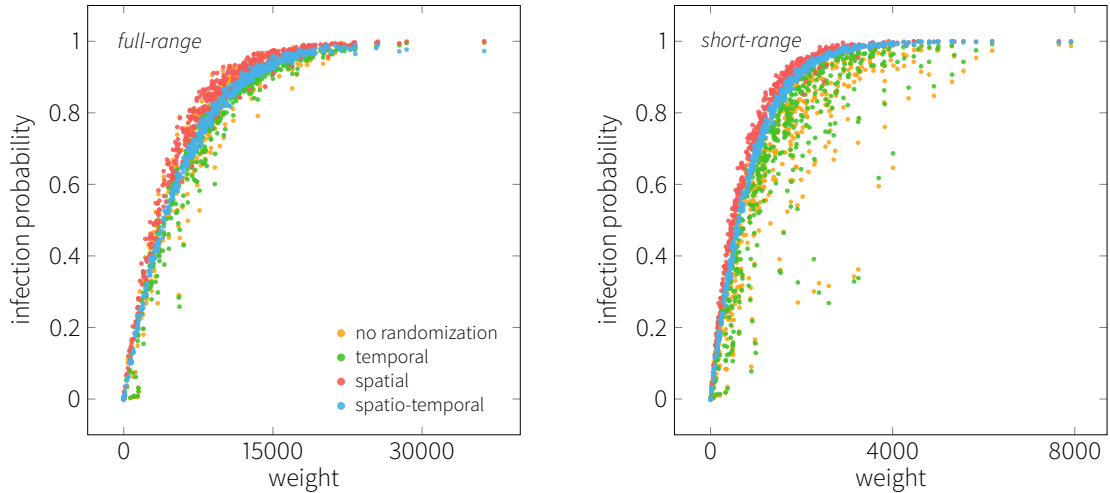

Figure 8: **Structural effects on monitoring performance** Infection probability of individuals versus their corresponding weight in the proximity networks after several correlations are removed: the original networks (yellow), after temporal correlations are removed (green), after structural correlations are removed (red) and when all correlations are eliminated (blue). In all cases, the personal weights are kept fixed as it forms the basis of the curves shown.

*Spatial* All degree-correlations are removed, but both circadian rhythm and the static degree-distributions are kept fixed. This corresponds to a redistribution of temporal links according to the degree-distributions and temporal edge density. Individual weights remain the same.

*Spatio-temporal* Both temporal and spatial correlations are removed, together with the aggregated degree-distribution. Only the individual weights are conserved from the original structure of the networks.

In both short- and full-range networks, removing spatio-temporal correlations also removes the vast majority of noise from the structure, and results in a clear trend of infection probability as a function of weight, saturating at some point (Figure 8). The higher level of initial noise in the original networks seen in the short-range case indicates that structure plays a major role in the strength of weight-based target groups. Spatial randomization also removes more noise from the trends, suggesting that the temporal nature of the networks has a smaller effect during epidemic surveillance. These observations point our that weight is the most relevant property when the susceptibility of individuals is considered. This is supported by the fact that highly central individuals in social networks are more likely to have high weight, we find the correlations between social network centralities and physical proximity networks to be approximately $r_{\text{Spearman}} = 0.4$. However, as we have seen, the structural correlation between social networks and proximity networks does not prove to be sufficient for efficient vaccination, as vaccination corresponds to a fundamentally different challenge, by effectively removing individuals from the network, thus changing the underlying network structure.

# S4  Vaccination

## S4.1  Social vaccination

In the manuscript we have shown that social vaccination is effective in short-range transmission, here we explore robustness of the observed performance against various definition of 'central individuals'. Several strategies exploiting the structure of social networks have been proposed for efficient immunization [10, 11, 9]. Figure 9a shows the performance of social vaccination (in short-range transmission) based on different strategies for determining target individuals. The network measures in the population are highly correlated and lead to qualitatively similar results, independent of whether degree, $k$-cores, or betweenness centrality are considered. We also examine different configurations of the social network channels used for social vaccination, adding text message metadata to call metadata and using the Facebook friendship graph. We find no significant difference in the efficacy by using these different views, indicating that the epidemiologically relevant individuals can be robustly identified based on the digital social networks.

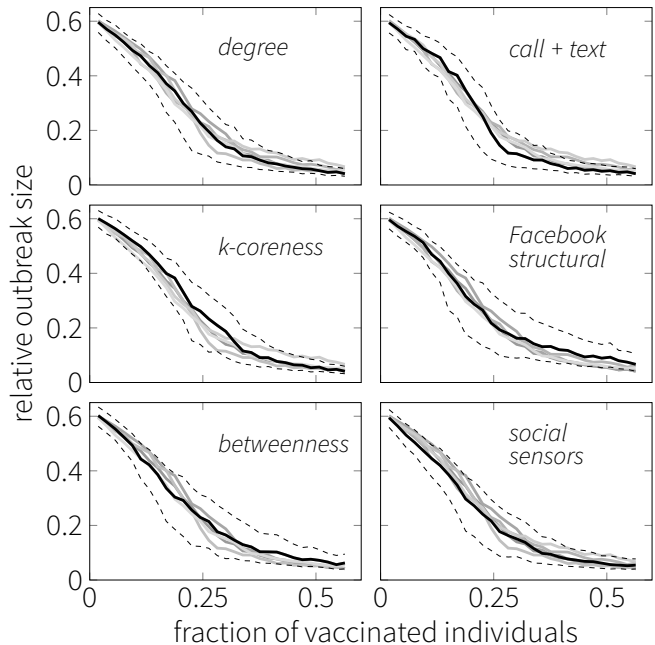

Figure 9: **Effect of various centrality measures and communication channels considered.** Panels show the vaccination performance of target groups based on different centralities and layers, with a single centrality or channel being highlighted in each panel. In case of different centralities (left panels), the call network was used as the basis for the analysis. For the different channels (right panels, with the exception of the bottom panel), we selected individuals based on the closeness centrality. Bottom right panel shows the results with the social sensors, the simulation of Christakis et al. social sensor selection method based on the call + text message data among students.

To overcome the difficulty of collecting a complete network of interactions for targeted interventions, a scheme of social sensors (acquaintance immunization) has been proposed [7, 8, 9]. By vaccinating friends of randomly selected individuals—as such friends tend to be more central in the social network—strictly *local* information can be used (majority of the above measures rely on global network information). Here we evaluate the performance of social sensors scheme directly, choosing three strongest contacts from the call and SMS lists of each individual as a proxy for naming someone a friend[1]. Individuals nominated as friends most often were then included in the target groups. We show that social sensors are a useful strategy for selecting individuals using strictly local information in the network, with the performance close to using full network structure (Fig. 9a lower right panel).

## S4.2 Models of social target groups

To understand the performance of social vaccination, here we provide a qualitative explanation of the observed vaccination curves. The performance of the social vaccination in the short- and full-range transmission networks can we be decomposed into the result of two basic mechanisms. For a given vaccination size, social networks manage to identify certain fraction of individuals from the optimal (colocation) group, the *the core*. Those target individuals not included in the colocation group (the *the periphery*) also contribute to the performance and in the case of an effective target group these individuals also display increased relevance during immunization. To understand the impact of these two components, we consider three modified models of target groups:

> *Model A* We keep the core fixed and replace the periphery with individuals sampled from outside the optimal group randomly.
>
> *Model B* We keep the periphery fixed and replace the core by randomly chosen individuals from the optimal group.
>
> *Model C* We keep the periphery fixed and replace the core with the same number of most connected participants from the optimal group.

Results are shown in Figure 10. The decrease of performance in model A indicates that social networks are able to locate more appropriate candidates (compared to random) outside of the optimal group in addition to the optimal ones (Fig. 10a). This observation is more pronounced in the short-range network (Fig. 10b). In case of the core however, social networks highlight optimal individuals with the same efficiency as one would obtain by sampling the optimal group randomly, as indicated by negligible change in performance in model B. Finally, the performance change of model C compared to the original strategy is higher in the full-range network than that in the short-range, indicating that this effect is a major driver (together with difference uncovered by model A) behind the superior performance of the social vaccination in the short-range network.

The difference in the performance due to the contribution of the periphery can be explained by the structural differences between the proximity networks. Calculating cosine similarity of the participants' weighted degree $v$ (the fraction of interactions they have with the rest of the population) between digital social and physical proximity networks for all participants $\theta(v_{\text{digital}}, v_{\text{physical}})$, we show that short-range network displays higher frequency of high similarity values ($\theta > 0.5$) (Fig. 10c). Digital social networks capture the local structure of the short-range network to a higher extent than that of the long-range network, which includes a high fraction of incidental interactions[3]. As a consequence, social networks are able to locate epidemiologically relevant target individuals outside of the optimal group, as central participants in the social networks are also exhibit high centrality in the short-range network.

On the other hand, performance change with respect to the core is related to whether the contacts of the vaccinated nodes are *internal* or *external*. Vaccinating individuals effectively removes them from the network along with all of their contacts. Internal links are connecting target individuals, and so their interactions do not contribute to the improvement in the vaccination. Thus, even though the total weight of the removed links may be equal in short- and full-range networks, the actual impact of the removed links can be different if the distribution of internal/external links differ. Interactions in the short-range network tend to take place between participants socially more related[3], and we observe that target groups in that network also tend to be more interconnected (Fig. 10d). Thus choosing the individuals with the highest level of interactions (weight) in the short-range network does not lead to pronounced increment in performance, as we remove individuals from a more interconnected subgraph (compared to the full-range network).

## S4.3 Epidemic spread on the social networks

The main aim of the work presented in the manuscript is to find individuals based on their social network (which is approximated from their digital fingerprint) while probing their impact in case of an epidemic outbreak that takes place on the physical proximity networks. Our results show that social and proximity networks are structured in a way that allows for a decreased performance for vaccination, however, the efficacy of state-of-the-art strategies proposed in the literature is related to the fact that both finding target individuals and evaluating their impact was performed on the same network. Figure 11b shows how we would understand the performance of targeted vaccination if both the selection and the spread of took place on the social networks. As we can see, if the disease spreads in the same network that is the basis for the target group selection, degree immunization (that is, one of the most efficient strategy) performs well in both dynamic and static cases, and significant performance change is observed compared to the, e.g., closeness centrality based strategy.

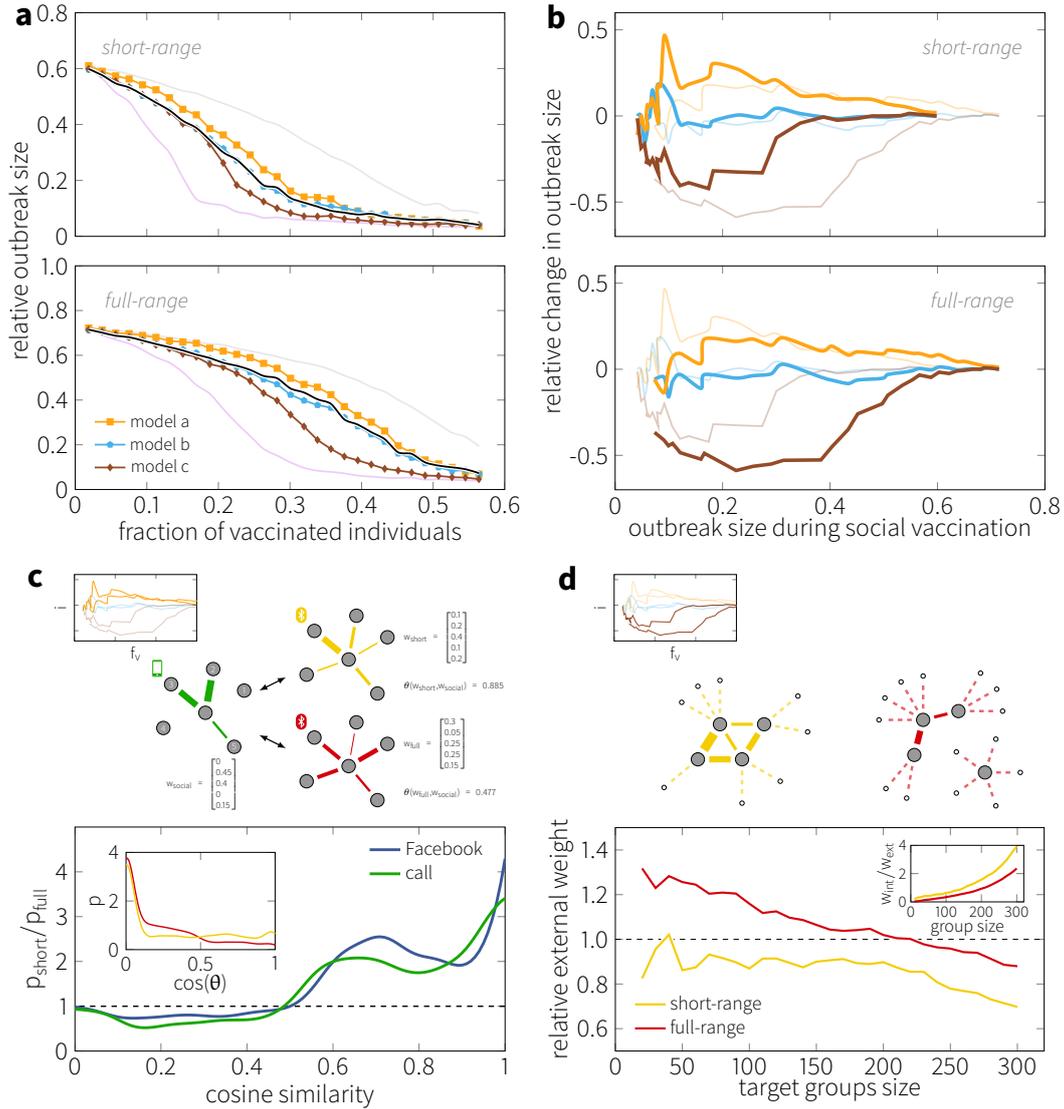

Figure 10: **Role of target individuals in social target groups. a)** Median relative outbreak size in three models (see text for their definitions) compared to random (light gray), optimal (light purple) and the call network (light green). **b)** Relative differences in the outbreak sizes of the original and the modified social groups as the function of the outbreak size in the optimal vaccination, the curves of the corresponding proximity network are highlighted. **c)** Top: illustration of the cosine similarity. For each individual, we define their weighted degree as the total fraction of interactions with all other participants (weighted degrees are normalized to have unit component sum). For a given proximity network, the cosine similarity between the weighted degrees of the same individual is calculated, and the distribution of similarities is constructed. Bottom: point-wise ratio of the probabilities of similarity values in short-range and full-range networks, when compared to the call (green) and Facebook (blue) graphs. Inset shows the original similarity distributions for the short-range (yellow) and full-range (red) graphs with the call network. **d)** Weight removed from the remainder of the graph after the removal of the vaccinated individuals. Top: illustration of the weights distributed among the removed (gray) and remaining (white) individuals. Solid lines correspond to internal, dashed denote external weights being removed from the network. Bottom: total external weight removed from the graphs when the overlap of the social and optimal target groups are replaced by the highest-weight (solid) or randomly chosen (dashed) individuals from the optimal group. Inset shows the ratio of internal / external weights in the two networks for different target groups.

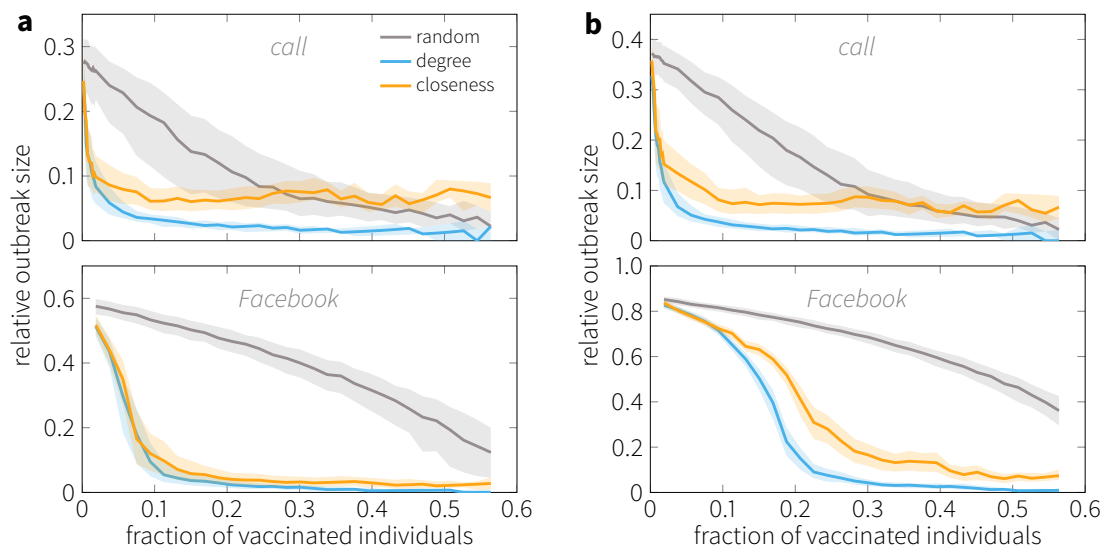

Figure 11: **Vaccination performance of outbreaks taking place on the social networks.** Efficient immunization strategies against epidemic that spreads via the ties in the digital social networks. In this case, both the target individuals are selected by and the epidemic takes place on the same social network. **a)** The network dynamics is also considered as they evolve in time, **b)** the static aggregated networks are used and the epidemic spreads rapidly on the static links.

## S5 Robustness

So far we have investigated the performance of the target group in the same period that formed the basis for their selection, however, in case of real epidemic scenarios it is preferred to collect data from a preceding periods so that we can infer to upcoming months. In this spirit, here we use the target groups of February (in the index month) for monitoring and vaccination in the consecutive months March, April and May. As an evaluation, we compare social target groups to the colocation based groups from both the index and later months.

Figure 12 shows the infection probability and infection time observed in the social target groups for different months, using the short-range proximity network. In the top panel, we show the infection probability relative to the probability measures in the optimal target groups based on the colocation data in the actual month. In all months, social target groups outperform random monitoring, and also show similar probability values to the optimal that can be obtained using proximity data of the same period. However, as the period in interest becomes further in time from the index month, efficacy becomes less stable and it decreases as well. Regarding infection time, we show the raw results in units of days compared to the optimal of the month. Again, social groups display significantly earlier times of infection than that of random groups but they are also less effective as the true optimal is. Interestingly, the decreasing performance trend observed in case of infection probability is not present here and social target groups perform equally well in all months.

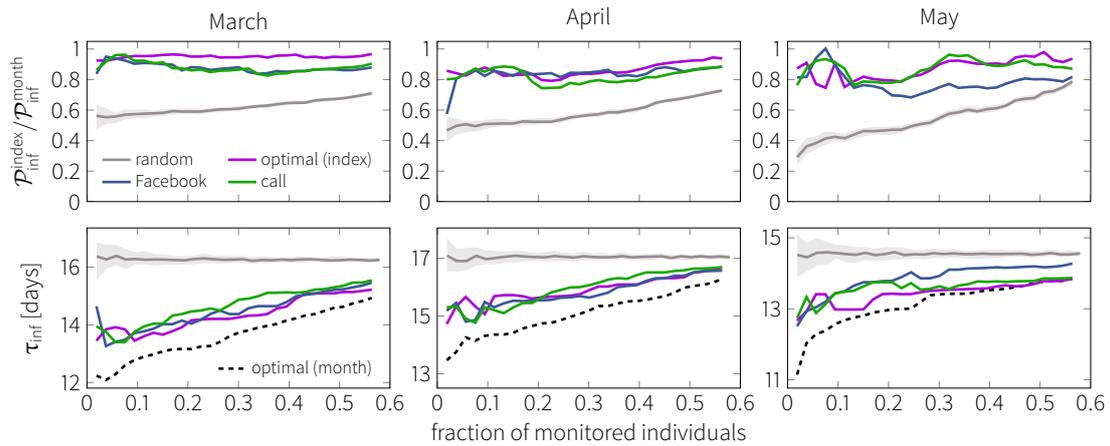

Figure 12: **Robustness of social target groups in external periods of time.** Top: infection probability relative to the current optimal. Bottom: infection time in target groups. Each curve shows the median over $10^4$ simulations with a minimum relative outbreak size of 20%. Error bars on the random group results denote lower and upper quartiles of 100 realizations.